\documentclass[english,aps,manuscript,aps,preprint]{revtex4}
\usepackage[T1]{fontenc}
\usepackage[latin9]{inputenc}
\usepackage{textcomp}
\usepackage{amsmath}
\usepackage{amssymb}
\usepackage{esint}

\makeatletter
\@ifundefined{textcolor}{}
{%
 \definecolor{BLACK}{gray}{0}
 \definecolor{WHITE}{gray}{1}
 \definecolor{RED}{rgb}{1,0,0}
 \definecolor{GREEN}{rgb}{0,1,0}
 \definecolor{BLUE}{rgb}{0,0,1}
 \definecolor{CYAN}{cmyk}{1,0,0,0}
 \definecolor{MAGENTA}{cmyk}{0,1,0,0}
 \definecolor{YELLOW}{cmyk}{0,0,1,0}
}

\makeatother

\usepackage{babel}
\begin{document}

\title{One-loop contribution to the matter-driven expansion of the Universe}

\author{Bogus\l aw Broda}

\email{bobroda@uni.lodz.pl}

\homepage{http://merlin.phys.uni.lodz.pl/BBroda/}

\affiliation{Department of Theoretical Physics, Faculty of Physics and Applied
Informatics, University of \L ód\'{z}, 90-236 \L ód\'{z}, Pomorska
149/153, Poland}
\begin{abstract}
Standard perturbative quantum gravity formalism is applied to compute
the lowest order corrections to the spatially flat cosmological FLRW
solution governed by ordinary matter. The presented approach is analogous
to the one used to compute quantum corrections to the Coulomb potential
in electrodynamics, or to the approach applied in computation of quantum
corrections to the Schwarzschild solution in gravity. In this framework,
it is shown that the corrections to the classical metric coming from
the one-loop graviton vacuum polarization (self-energy) have (UV cutoff
dependent) repulsive properties, which could be not negligible in
the very early Universe.


\textbf{PACS (2010) numbers: }04.60.Gw Covariant and sum-over-histories
quantization; 98.80.Es Observational cosmology (including Hubble constant,
distance scale, cosmological constant, early Universe, etc): 04.60.Bc
Phenomenology of quantum gravity.
\end{abstract}
\maketitle

\paragraph*{Introduction.}

The aim of our work it to explicitly show the appearance of ``repulsive
forces'' of quantum origin, which could be not negligible in the
very early evolution of the Universe. Actually, we apply the method
used earlier in the case of the spatially flat Friedmann--Lemaître--Robertson--Walker
(FLRW) solution governed by radiation \cite{broda2011one} (see also
\cite{broda2011quantum,broda2014perturbative}). It appears that the
cosmological FLRW case with ordinary matter as a source is similar
to the radiation one. Namely, the lowest order quantum corrections
coming from the fluctuating graviton vacuum are ``repulsive'', resembling
the situation well-known in loop quantum cosmology (LQC) \cite{ashtekar2009loop,Ashtekar:2003hd,Bojowald:2006da}.
The phenomenon is obviously negligible in our epoch, but it could
be not so in the very early Universe. One should stress that the derivation
is a lowest order approximation---the graviton vacuum polarization
(self-energy) is taken in one-loop approximation, and the approach
assumes the validity of the weak-field regime.

\paragraph*{One-loop corrections.}

Our starting point is a general spatially flat FLRW space-time
\begin{equation}
ds^{2}=g_{\mu\nu}dx^{\mu}dx^{\nu}=dt^{2}-a^{2}(t)\boldsymbol{dr}^{2},\label{eq:general FLRW metric}
\end{equation}
with the cosmic time-dependent scale factor $a(t)$. To satisfy the
condition of weakness of the gravitational field $\kappa h_{\mu\nu}$
near the reference time $t=t_{0}$ in the perturbative expansion
\begin{equation}
g_{\mu\nu}=\eta_{\mu\nu}+\kappa h_{\mu\nu},
\end{equation}
($\kappa=\sqrt{32\pi G_{N}}$, with the Newtonian gravitational constant
$G_{N}$), the metric is rescaled in such a way that it is exactly
Minkowski one for $t=t_{0}$, i.e.
\begin{equation}
a^{2}(t)=1-\kappa h(t),\qquad h(t_{0})=0.\label{eq:metric}
\end{equation}
Then
\begin{equation}
h_{\mu\nu}(t,\boldsymbol{r})=h(t)\mathcal{I_{\mu\nu}}\quad\mathrm{and}\quad\mathcal{I}_{\mu\nu}\equiv\left(\begin{array}{cc}
0 & 0\\
0 & \delta_{ij}
\end{array}\right).\label{eq:h mi ni eq. h I mi ni}
\end{equation}
In view of the standard harmonic gauge condition (see the second eq.\ in
(\ref{eq:our barred metric})), which we impose, we perform the following
gauge transformation:
\begin{equation}
\kappa h_{\mu\nu}\rightarrow\kappa h_{\mu\nu}^{'}=\kappa h_{\mu\nu}+\partial_{\mu}\xi_{\nu}+\partial_{\nu}\xi_{\mu}\quad\textrm{with }\quad\xi_{\mu}\left(t\right)=\left(-\frac{3\kappa}{2}\int_{0}^{t}h(t')\,dt',\,0,\,0,\,0\right).\label{eq:gauge transformation}
\end{equation}
For simplicity, skipping the prime, we get
\begin{equation}
h_{\mu\nu}(t,\boldsymbol{r})=h(t)\left(\begin{array}{cc}
-3 & 0\\
0 & \delta_{ij}
\end{array}\right)\quad\textrm{and }\quad h{}_{\lambda}^{\lambda}(t)=-6h(t),\label{eq:new metric tensor}
\end{equation}
where the indices are being manipulated with the flat Minkowski metric
$\eta_{\mu\nu}$. Switching from $h_{\mu\nu}$ to standard (``better'')
perturbative gravitational variables, namely to the ``barred'' field
$\bar{h}_{\mu\nu}$ which is defined by
\begin{equation}
\bar{h}_{\mu\nu}\equiv h_{\mu\nu}-\tfrac{1}{2}\eta_{\mu\nu}h_{\lambda}^{\lambda},\label{eq:general barred metric}
\end{equation}
we get 
\begin{equation}
\bar{h}_{\mu\nu}(t,\boldsymbol{r})=-2h(t)\mathcal{I}_{\mu\nu}\quad\textrm{with}\quad\partial^{\mu}\bar{h}_{\mu\nu}=0.\label{eq:our barred metric}
\end{equation}
The Fourier transform of $\bar{h}_{\mu\nu}$ is of the form 
\begin{equation}
\tilde{\bar{h}}_{\mu\nu}(p)=-2\tilde{h}(E)\left(2\pi\right)^{3}\delta^{3}(\boldsymbol{p})\mathcal{I}_{\mu\nu}.\label{eq:fourier metric}
\end{equation}

To obtain quantum corrections to classical field we should supplement
the classical line with a vacuum polarization (self-energy) contribution
and a corresponding (full) propagator. Therefore, the lowest order
quantum corrections $\tilde{\bar{h^{\textrm{q}}}}_{\mu\nu}$ to the
classical gravitational field $\tilde{\bar{h^{\textrm{c}}}}_{\mu\nu}$
are given, in the momentum representation, by the formula (see, e.g.\ \cite{QuantumcorrectionstotheSchwarzschildsolution},
or §~114 in \cite{QuantumElectrodynamics} for an electrodynamic
version---the so-called Uehling potential)
\begin{equation}
\tilde{\bar{h^{\textrm{q}}}}_{\mu\nu}(p)=\left(D\Pi\tilde{\bar{h^{\textrm{c}}}}\right)_{\mu\nu}(p),\label{eq:quantum from classical 1}
\end{equation}
where
\begin{equation}
D_{\mu\nu}^{\alpha\beta}(p)=\frac{i}{p^{2}}\mathbb{D_{\mu\nu}^{\alpha\beta}}\label{eq:propagator}
\end{equation}
is the free graviton propagator in the harmonic gauge with the auxiliary
(constant) tensor $\mathbb{D}$ defined below in Eq.(\ref{eq:D,E,P}),
and $\Pi_{\mu\nu}^{\alpha\beta}(p)$ is the (one-loop) graviton vacuum
polarization (self-energy) tensor operator. Now, we define the auxiliary
tensors:
\begin{equation}
\mathbb{D}\equiv\mathbb{E}-2\mathbb{P},\quad\textrm{where}\quad\mathbb{E}_{\mu\nu}^{\alpha\beta}\equiv\tfrac{1}{2}\left(\delta_{\mu}^{\alpha}\delta_{\nu}^{\beta}+\delta_{\nu}^{\alpha}\delta_{\mu}^{\beta}\right)\quad\textrm{and}\quad\mathbb{P}_{\mu\nu}^{\alpha\beta}\equiv\tfrac{1}{4}\eta^{\alpha\beta}\eta_{\mu\nu};\label{eq:D,E,P}
\end{equation}
which satisfy the following useful identities:
\begin{equation}
\mathbb{E}^{2}=\mathbb{E},\quad\mathbb{P}^{2}=\mathbb{P},\quad\mathbb{EP}=\mathbb{PE}=\mathbb{P}\quad\textrm{and}\quad\mathbb{D}^{2}=\mathbb{E}.\label{eq:E, P, D identities}
\end{equation}
By virtue of the definition (\ref{eq:general barred metric}), we
find that 
\begin{equation}
\bar{h}_{\mu\nu}=\left(\mathbb{D}h\right)_{\mu\nu}.\label{eq:h bar eq Dh}
\end{equation}
Multiplying Eq.(\ref{eq:quantum from classical 1}) from the left
by $\mathbb{D}$, we get (using (\ref{eq:propagator}), (\ref{eq:h bar eq Dh}),
and the last identity in (\ref{eq:E, P, D identities}))
\begin{equation}
\tilde{h^{\textrm{q}}}_{\mu\nu}(p)=\frac{i}{p^{2}}\left(\Pi\tilde{\bar{h^{\textrm{c}}}}\right)_{\mu\nu}(p).\label{eq:quantum from classical}
\end{equation}
Actually, a substantial simplification takes place in Eq.(\ref{eq:quantum from classical}),
namely,
\begin{equation}
\tilde{h^{\textrm{q}}}_{\mu\nu}\left(p\right)=\frac{i}{p^{2}}\left(\Pi'\tilde{\bar{h^{\textrm{c}}}}\right)_{\mu\nu}\left(p\right),\label{eq:quantum from classical prime}
\end{equation}
where $\Pi'(p)$ is an ``essential'' part of the full (in the sense
of the one-loop approximation) graviton polarization operator $\Pi(p)$.
The ``essential'' part $\Pi'\left(p\right)$ of the full (one-loop)
graviton vacuum polarization operator $\Pi\left(p\right)$ can obtained
from $\Pi\left(p\right)$ by skipping all the terms with the momenta
$p$ with free indices (e.g. $\alpha$, $\beta$, $\mu$, or $\nu$).
Such a simplification follows from the gauge freedom the $\tilde{h^{\textrm{q}}}_{\mu\nu}$
enjoys, and from the harmonic gauge condition the ${\tilde{\bar{h^{\textrm{c}}}}}_{\alpha\beta}$
should satisfy. In general, by virtue of the symmetry of the indices,
$\Pi\left(p\right)$ consists of 5 (tensor) terms. Each $p_{\mu}$
can be ignored in $\Pi\left(p\right)$ because it only gives rise
to a gauge transformation of $\tilde{h}_{\mu\nu}$. Moreover, since
$\bar{\tilde{h}}_{\alpha\beta}$ satisfies the harmonic gauge condition,
the terms with $p^{\alpha}$ in $\Pi\left(p\right)$ are annihilated.
In other words, schematically
\begin{equation}
\Pi\left(p\right)=\Pi'\left(p\right)+\;\cdots p\cdots\;.
\end{equation}
Since the momenta $p$ in the ellipses posses free indices, they can
be ignored, and only the first two terms with dummy indices (i.e.\ $p^{2}$)
survive, i.e.
\begin{equation}
\Pi'(p)=\kappa^{2}p^{4}I(p^{2})(2\alpha_{1}\mathbb{E}+4\alpha_{2}\mathbb{P}),\label{eq:Pi'}
\end{equation}
where the numerical values of the coefficients $\alpha_{1}$ and $\alpha_{2}$
depend on the kind of the field circulating in the loop, and the (scalar)
standard loop integral $I(p^{2})$ with the UV cutoff denoted by $M$
is asymptotically of the form (see, e.g., Chapt.\ 9.4.2 in \cite{QuarksLeptonsandGaugeFields})
\begin{equation}
I(p^{2})=\frac{1}{\left(2\pi\right)^{4}}\int\frac{d^{4}q}{q^{2}\left(p-q\right)^{2}}=-\frac{i}{\left(4\pi\right)^{2}}\log\left(-\frac{p^{2}}{M^{2}}\right)+\cdots\quad,\label{eq:asymptotic integral}
\end{equation}
where the dots mean terms $\mathcal{O}\left(p^{2}/M^{2}\right)$.
A standard way to derive (\ref{eq:asymptotic integral}) consists
in continuing from $q_{0}$ to $+iq_{4}$ ($d^{4}q\rightarrow id^{4}q_{\mathrm{E}})$,
exponentiating the denominator using a (double) proper-time representation
for the propagators, a change of proper-time variables, imposing the
UV cutoff for a new proper time, and next continuing back to the Minkowski
momentum variables. Thus, we obtain

\[
\tilde{h^{\textrm{q}}}_{\mu\nu}(p)=\frac{i}{p^{2}}\kappa^{2}p^{4}\left[-\frac{i}{\left(4\pi\right)^{2}}\log\left(-\frac{p^{2}}{M^{2}}\right)\right]\left[-2\tilde{h^{\mathrm{c}}}(E)\left(2\pi\right)^{3}\delta^{3}(\boldsymbol{p})\right]\left[(2\alpha_{1}\mathbb{E}+4\alpha_{2}\mathbb{P})\mathcal{I}\right]_{\mu\nu}
\]
\begin{equation}
=-2\pi\kappa^{2}E^{2}\log\left|\frac{E}{M}\right|\tilde{h^{\mathrm{c}}}(E)\delta^{3}(\boldsymbol{p})\left(\begin{array}{cc}
-3\alpha_{2} & 0\\
0 & \left(2\alpha_{1}+3\alpha_{2}\right)\delta_{ij}
\end{array}\right).\label{eq:quantum h(t)}
\end{equation}

\paragraph*{Matter source.}

Now, we specify our input classical metric. To this end, we choose
the matter source assuming
\begin{equation}
a^{2}(t)=\left|\frac{t}{t_{0}}\right|^{4/3}.\label{eq:m-radiation}
\end{equation}
According to (\ref{eq:metric}) and (\ref{eq:m-radiation}) the Fourier
transform of $h^{\mathrm{c}}(t)$ is
\begin{equation}
\tilde{h^{\mathrm{c}}}(E)=\frac{2}{\kappa t_{0}^{4/3}}\sin\left(2\pi/3\right)\Gamma\left(7/3\right)\left|E\right|^{-7/3}+\cdots,\label{eq:Fourier(h)}
\end{equation}
where the dots mean a term (vanishing in (\ref{eq:quantum h(t)}))
proportional to the Dirac delta. Performing the gauge transformation
in the spirit of (\ref{eq:gauge transformation}), we remove the purely
time component of ${h^{\mathrm{q}}}_{\mu\nu}$, i.e.\ ${h^{\mathrm{q}}}_{00}\rightarrow{h^{\mathrm{q}}}'_{00}=0$.
The inverse Fourier transform yields now the quantum correction
\begin{equation}
{h^{\mathrm{q}}}_{\mu\nu}(t)=\frac{\alpha\kappa}{\left(3\pi\right)^{2}t_{0}^{4/3}}|t|^{-2/3}\left(\log\left|t/t_{c}\right|+c\right)\mathcal{I}_{\mu\nu},\label{eq:hqmn(t)}
\end{equation}
where $c\equiv\gamma+\frac{3}{2}\log3+\frac{\pi}{3}\sqrt{3}$ ($\gamma$
is  the Euler--Mascheroni constant), $t_{c}$ is an UV cutoff in time
units, and $\alpha\equiv2\alpha_{1}+3\alpha_{2}$. According to the
table given in \cite{broda2011one} only the graviton field yields
a non-zero contributions with $\alpha=-\tfrac{1}{16}$. Now one can
easily check that the second time derivative of (\ref{eq:hqmn(t)})
is positive for $t>t_{c}$. Therefore the quantum contribution to
the classical expansion is accelerating.

\paragraph*{Final remarks.}

In the framework of the standard (one-loop) perturbative quantum gravity,
we have derived the formula (\ref{eq:hqmn(t)}) expressing a contribution
to the classical metric governing matter-driven expansion of the Universe.
As the second time derivative of (\ref{eq:hqmn(t)}) is positive we
expect an accelerating role of this contribution, especially in early
evolution of the Universe. In spite of the fact that (\ref{eq:hqmn(t)})
is UV cutoff dependent, the (qualitative) result is unchanged as long
as $t>t_{c}$. 

Supported by the University of \L ód\'{z} grant.

\bibliographystyle{apsrev}
\bibliography{One-loop_contribution_to_the_matter-driven_expansion_of_the_Universe}

\end{document}